\documentstyle[psfig]{basi}
\begin{document}
\title[Seyfert galaxies: Nuclear Radio 
        Structure and Unification]{Seyfert Galaxies: Nuclear Radio 
        Structure and Unification}
\author[D.V. Lal]%
       {Dharam Vir Lal\thanks{e-mail: dharam@iiap.ernet.in} \\ 
        Indian Institute of Astrophysics, Bangalore 560 034 \\
        Joint Astronomy Programme, Dept of Physics, 
        Indian Institute of Science, Bangalore 560 012}
\maketitle
\label{firstpage}
\begin{abstract}
We have observed a sample of Seyfert (Sy) galaxies with the
global Very Long Baseline Interferometry (VLBI)
to test the predictions of the unification scheme for
the Sy galaxies which hypothesises that Sys of Type~1 and of Type~2
constitute a single population of objects. If Sy~1 and Sy~2 galaxies
differ only in the orientation of the axisymmmetric active nucleus with respect
to the observer, then, the parsec~scale radio structures of the two types
should be similar.

The overall goal of the thesis is to rigorously test the predictions of the
unified scheme hypothesis for the Sy galaxies.

We chose a sample of 10 Sy~1s and 10 Sy~2s such that the two subsamples
were intrinsically similar within the framework of the unified scheme. This
was done by matching them as far as possible in the orientation-independent
parameters {\it viz.}, host galaxy properties and measures of intrinsic Active
Galactic Nuclei (AGN) power.

We here present the results obtained from these observations and their
implications on the unification scheme. We also discuss the results
obtained from the measurements at all IRAS and 2--10~keV X-ray wavebands
for our Sy galaxy sample using data from the literature.
\end{abstract}

\section{Introduction}
\label{sec:intro}
Sy galaxies are nearby, low luminosity AGNs which 
occur mostly in spiral hosts and are usually
taken to be ``radio-quiet" objects.
There are two kinds of Sy~galaxies, types~1 and 2, distinguished
by the widths of their spectrophotometrically observable emission
lines; the implied kinematic Doppler widths are ~$>$~1000~km~s$^{-1}$
and $<$ 1000 km~s$^{-1}$ for the two types, respectively.
Accordingly, we define a Sy~galaxy as a low-luminosity
(M$_{B} > -23.0$), radio-quiet (ratio of 5~GHz to {\it B}--band flux
density $<$ 10) object whose host galaxy~is a spiral, with Sy~1
galaxies having H$\beta$$_{\rm FWHM}$ (or H$\alpha$$_{\rm FWHM}$)
nuclear line widths exceeding 1,000 km~s$^{-1}$
(Khachikian \& Weedman 1974), and Sy~2 galaxies having~the ratio
of the [O~{\small III}]~$\lambda 5007$ to H$\beta$ line intensities
exceeding 3 (Dahari \& De~Robertis 1988).

The unified scheme for Sys hypothesises that Sy
of type~1 and 2 comprise a single population and appear different due to
the orientation of the axisymmetric active nucleus with respect to the
observer. According to the unification scheme, the torus of 
obscuring material is present
between the broad line region and the observer in all Sys.
The Sy~1s are those where we have a direct view of the central
engine and the Sy~2s are those where our line of sight to the central
engine is blocked by the torus. Broad emission lines ({\it i.e.}
with implied Doppler widths ~$>$~1000~km~s$^{-1}$) have been detected in
a few Sy~2s in {\it polarised} light (Antonucci \& Miller 1985). This
result strongly supports the unification scheme hypothesis,
because the spectrum of the polarised
flux in these Sy~2s ({\it e.g.} NGC~1068) showed permitted broad
emission lines of hydrogen and Fe~{\small {II}} and was thus very similar
to the spectrum of the total flux in Sy~1s.

In the following section \ref {the_out} we discuss, in brief,
the organisation of the thesis.

\subsection{Outline of the thesis}
\label {the_out}

The Chapter~1 of the thesis introduces `Active galaxy'
and  the various types of active galactic nuclei
that are generally recognised. We then focus on Sy
galaxies, develop the unification model for Sy~1 and Sy~2 galaxies. 
We also present several pieces of evidence in support and against
this model.

In Chapter 2, the result that poses a challenge to the
unified scheme (Section \ref {goal}), which we aim to address,
is discussed.
We introduce aperture synthesis and 
VLBI which is the key to achieve parsec~scale
resolution. We
then outline our rationale for the selection of our sample and
describe its construction (Section~\ref {samp_cons}).

In Chapter~3, we first give the
description of the observing procedure, data reductions and
analysis.
We give the description of the images for our sample sources
supplemented by the data from the
literature wheresoever necessary.

Chapter~4 uses results from Chapter~3 within the formalism
in which our sample was constructed to test the unified scheme.
We present our interpretations in the framework of
the starburst {\it versus}
accretion powered central engine debate as well as the unified scheme
hypothesis and relativistic beaming.

Chapter~5 compares the radio data with the data collated at other
wavebands that are published in the literature. We primarily
focus on the infrared and the X-ray emission.

The final chapter (Chapter~6) summarises the main results obtained
in the thesis and brings together the conclusions from this work,
highlighting our contribution to the understanding of Sy
galaxy phenomenology. We also critically assess the strengths and
weaknesses of our work and highlight opportunities for further
study.

\section{Aim of the research}
\label {goal}

Sy~galaxies have low luminosity radio emission, but they do show radio
emitting jet-like structures on small scales which appear to be the
low-power analogues of jets seen in radio powerful AGNs ({\it e.g.}
Nagar {\it et~al.} 1999). The unified scheme predicts that 
the total radio emission
should be similar in the two classes~of Sys (since the radio emission
is unattenuated by the obscuring torus), and their radio structures should
differ only due to the projection effects. However, this issue is controversial;
{\it e.g.} Roy {\it et~al.} (1994), using the 275--km long single baseline
Parkes-Tidbinbilla interferometer at 2.3~GHz, reported that Sy~2s show
more compact radio emission than Sy~1s. This result is inconsistent
with the predictions of~the unification scheme hypothesis.
The inconsistency remains even if
mild relativistic beaming is invoked, because in this case, the face-on
AGNs, {\it viz.}, Sy~1s, would be more likely to show compact
structures. {\it Our goal was to test predictions of the unified scheme, by
investigating the parsec~scale radio morphology of Sys using a
matched sample of Sy~1 and Sy~2 galaxies}.

\section{The sample}
\label {samp_cons}

Our sample selection criteria were as follows:
(i) the object should be a {\it bona~fide} Sy~galaxy ({\it cf.}~our
definition),
(ii)~it must be in a host galaxy that is a confirmed spiral,
(iii) it must be detected with $\sim$1 arcsecond resolution at 5--8~GHz
and~have~a detected compact component brighter than ~8~mJy at these
frequencies~and~on~these scales ({\it i.e.} as observed by VLA~$A$~or~$B$ array;
this criterion was required to make our experiment feasible), and
(iv) the host galaxy must have observed ratio of minor and major
isophotal diameter axes~$>$~0.5; we thereby exclude edge-on host galaxies
so as to minimise selection effects due to obscuration.
We note that Pringle {\it et al.} (1998) and Nagar \& Wilson (1999)
have shown that there is no correlation between the host galaxy
rotation axis and the direction of the radio jet.

\smallskip
From all Sys with available nuclear radio flux density at
$\sim$1~arcsecond resolution in the literature ({\it i.e.} all VLA~$A$,
\& $B$ array observations of Sys) we chose 10 Sy~1s and 10 Sy~2s
meeting the above criteria, such that the two sub-samples had similar
distributions of heliocentric redshift, luminosity of the host galaxy
({\it i.e.} minus the AGN) in the optical {\it B}--band,
[O~{\small III}]~$\lambda 5007$~luminosity, and galaxy bulge
luminosity. Thereby we ensured
that the sub-samples of Sy~1s and Sy~2s are {\it matched}, as far as
possible, with respect to their intrinsic AGN power and host galaxy
properties using {\it orientation-independent} parameters.

\section{Observations}

We observed 15 objects from our sample in Feb 1998 at 5~GHz using a
14--station global VLBI array,
including the phased VLA. We thus have
simultaneous VLA data also (angular resolution
$\sim$~$1^{\prime\prime}.0$) for all these objects. Of the remaining
5 sample objects, 4 have VLBI data 
in the literature which we add
to our own data and use in inferring our results below.

\subsection{Aperture synthesis and VLBI}

Aperture synthesis is a method where two or more antennas are
arranged so that their relative
positions may be altered to occupy successively the whole area of
a much larger equivalent antenna. By combining mathematically the
information derived from these different positions, it is
possible to obtain high resolution maps equivalent to that of a
single large antenna.

Interferometry is the pre-eminent method of achieving high angular
resolution. Long baseline interferometry achieves high angular
resolution by using two or more widely ($>$~1,000~km) separated
radio telescopes. VLBI was born in the
mid 1960's out of the wish of astronomers to achieve better
angular resolution than a regular radio interferometer.
It is a special case of the aperture
synthesis and is an extension of interferometric
techniques to the largest obtainable baselines. With telescopes
separated by intercontinental distances and operating
at cm~wavelengths,
a resolution of about 1~milliarcsec~scale can be achieved.

\section{Results}

\begin{figure}
\begin{center}
\hbox{
\psfig{figure=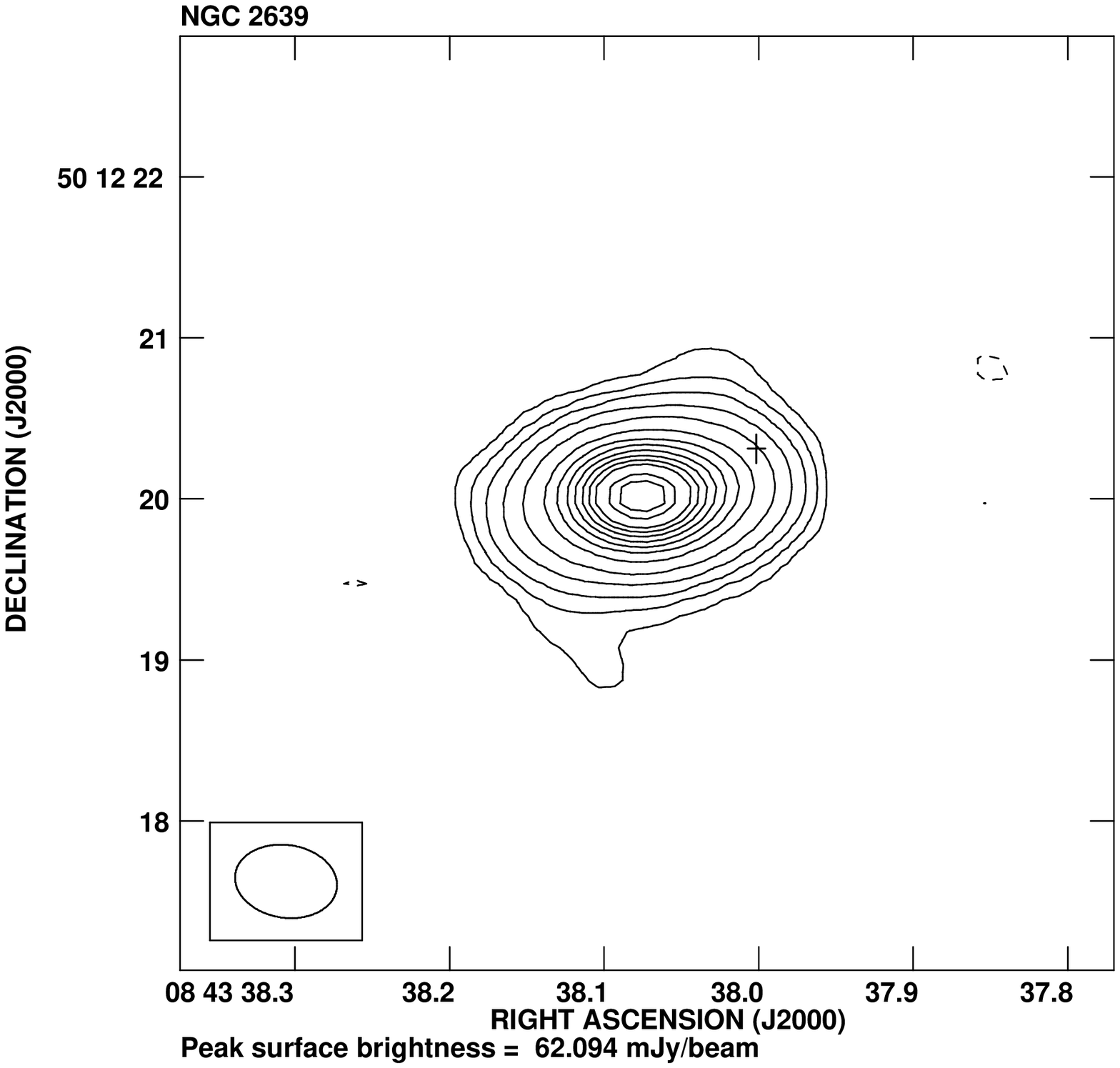,width=2.2in}
\hspace{1.0cm}
\psfig{figure=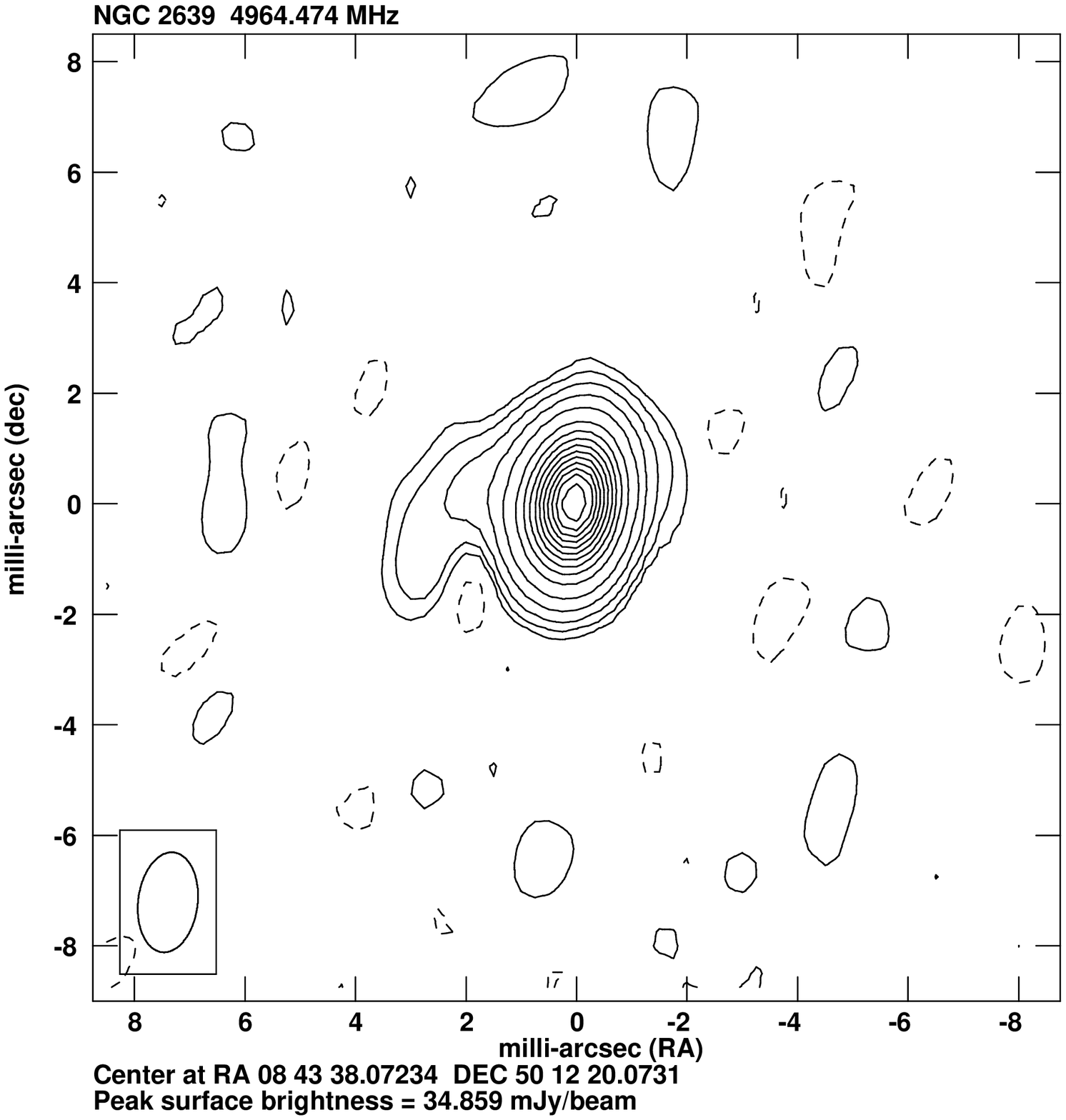,height=2.2in} }
\hbox{
\psfig{figure=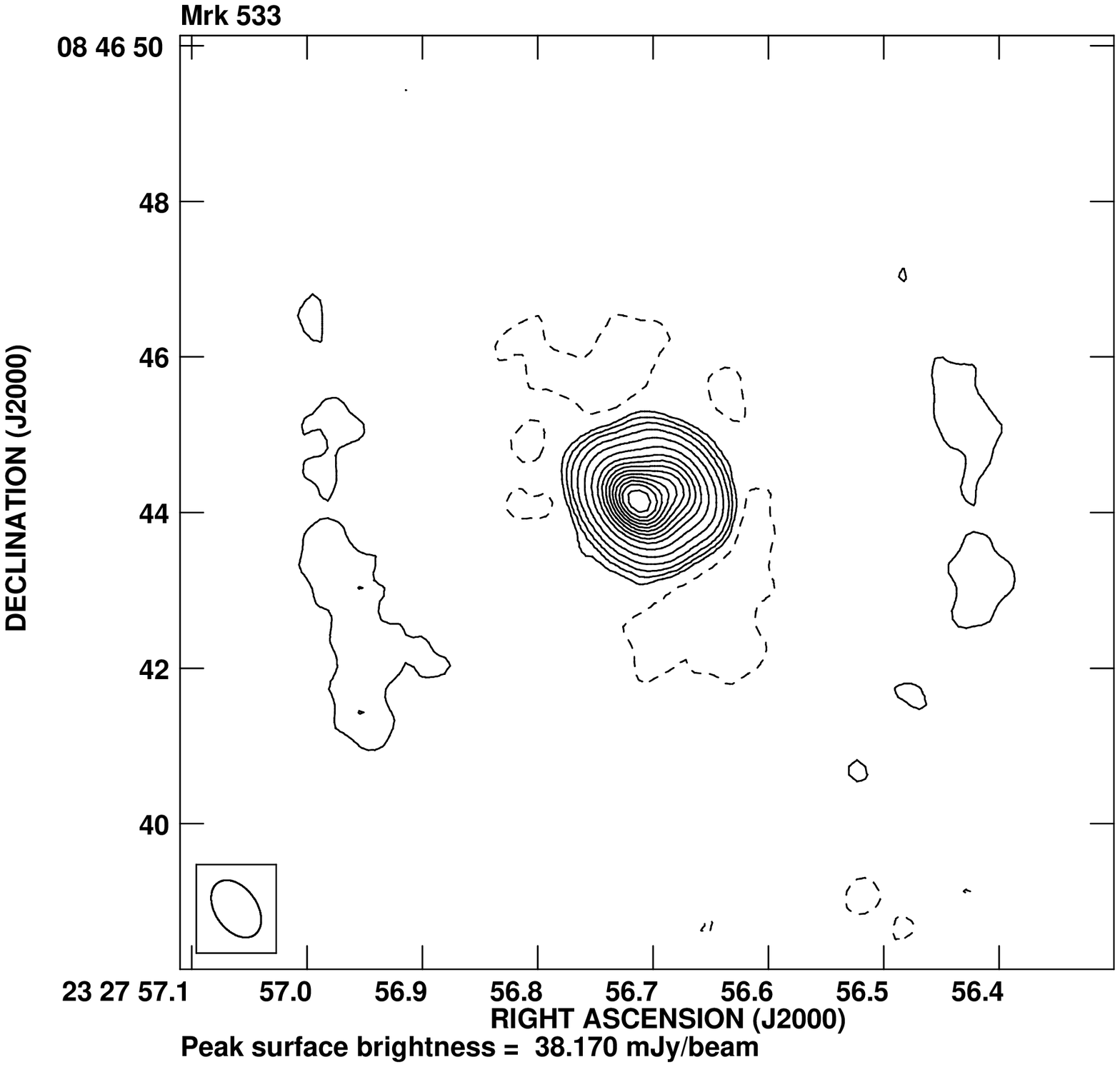,width=2.2in}
\hspace{1.0cm}
\psfig{figure=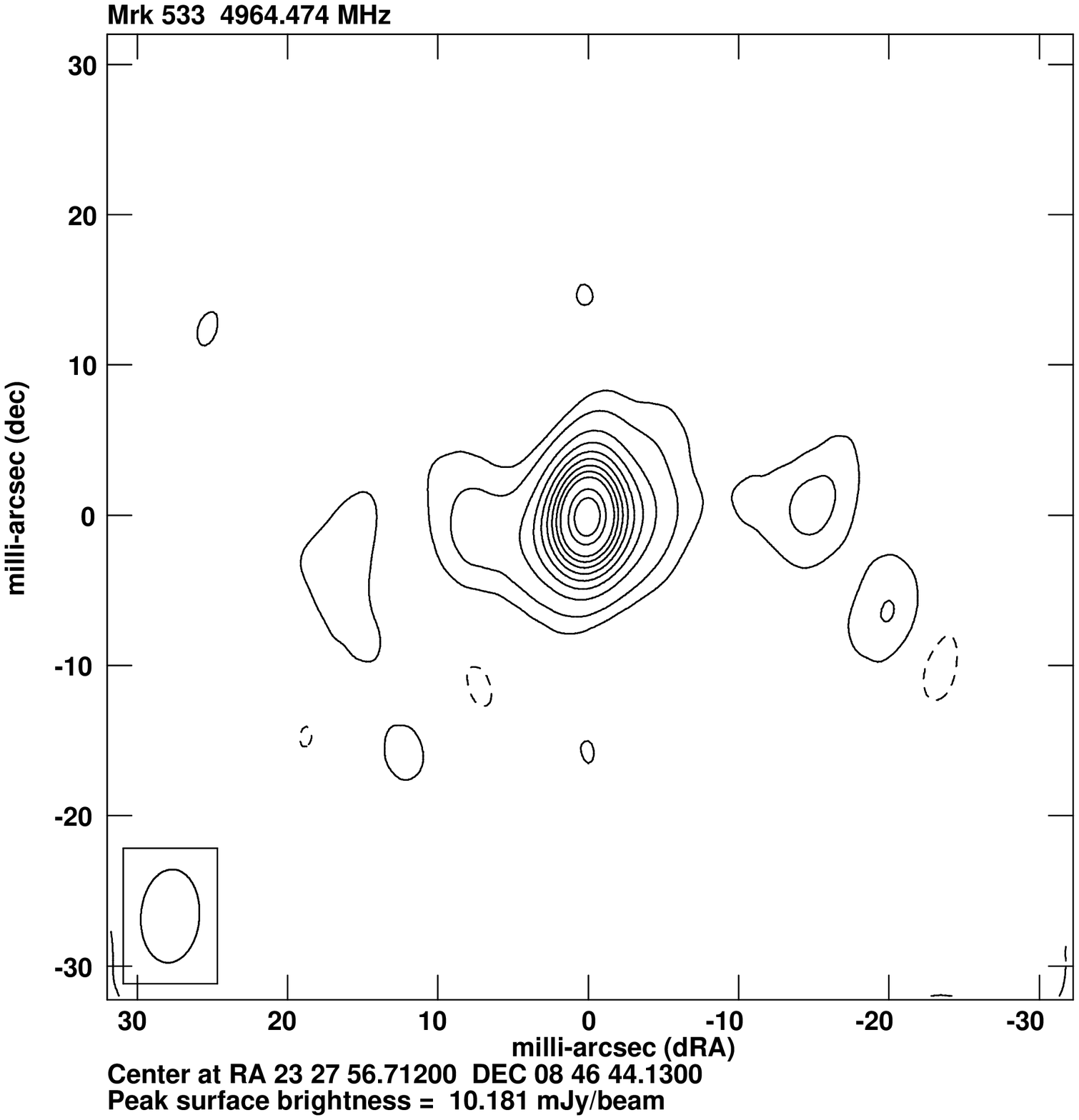,height=2.2in} }
\end{center}
\caption{\footnotesize
{6~cm {\small VLA} ({\it Upper, left panel}; the contour levels are at
--0.4, 0.4, 1, 2, 4, 8, 16, 24, 32, 40, 48, 56, 64, 80 \& 90\% of the peak
surface brightness) \& {\small VLBI}
({\it Upper, right panel}; the contour levels are at --0.5, 0.5, 1, 2,
4, 8, 16, 24, 32, 40, 48, 56, 64, 72, 80 \& 90\% of the peak
surface brightness) image of NGC~2639 and}
{6~cm {\small VLA} ({\it Lower, left panel}; the contour levels are at
--0.4, 0.4, 1, 2, 4, 8, 15, 24, 32, 40, 48, 56, 64, 72, 80 \& 90\% of the peak
surface brightness) \& {\small VLBI}
({\it Lower, right panel}; the contour levels are at --4, 4, 8, 16, 24, 32,
40, 48, 56, 64, 80 \& 90\% of the peak
surface brightness) image of Mrk~533. The small ellipse at the
corner of the image gives the size ({\small FWHM}) of the beam.} }
\label{fig1}
\end{figure}

\begin{figure}
\begin{center}
\hbox{
\psfig{figure=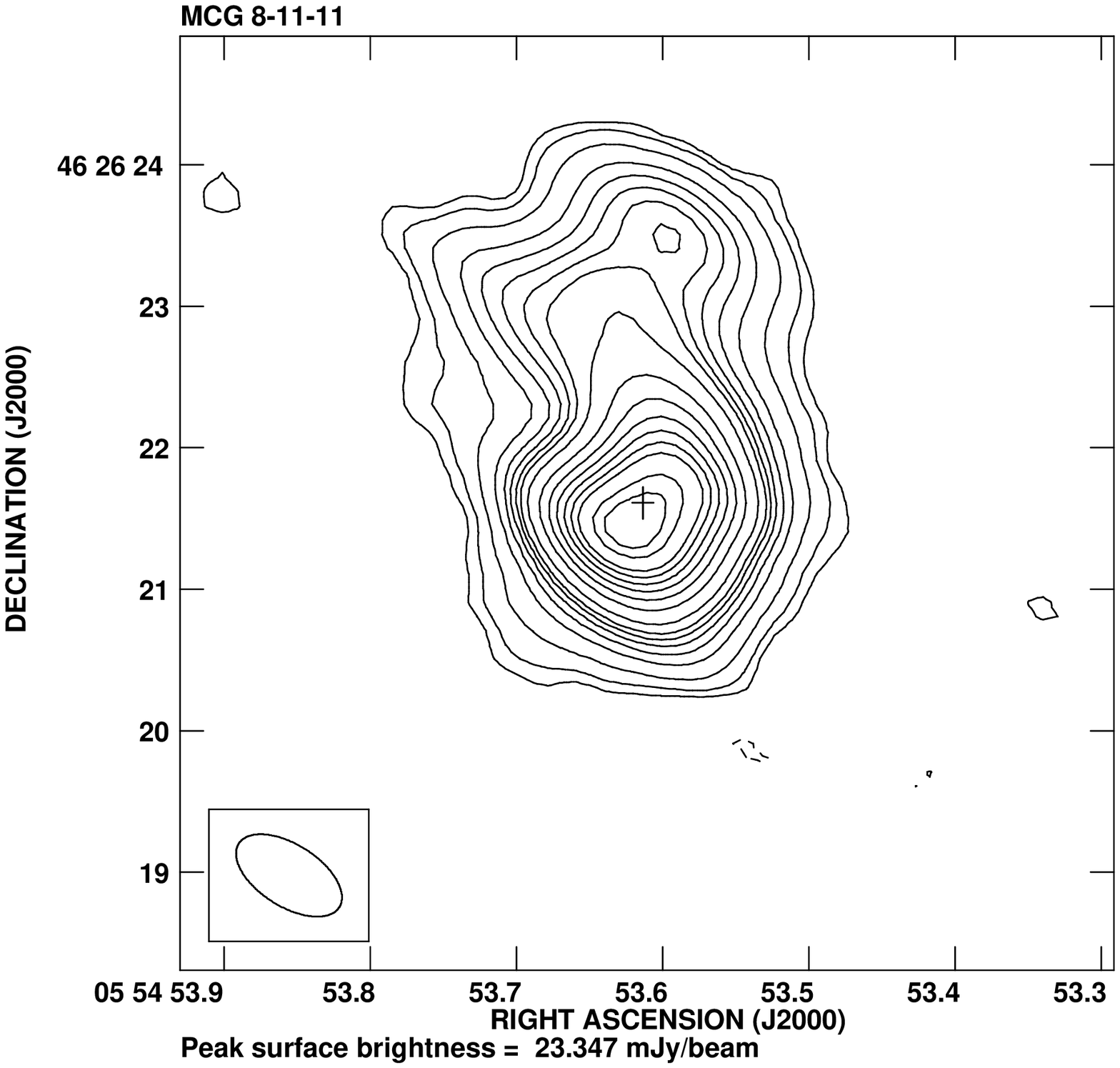,width=2.2in}
\hspace{1.0cm}
\psfig{figure=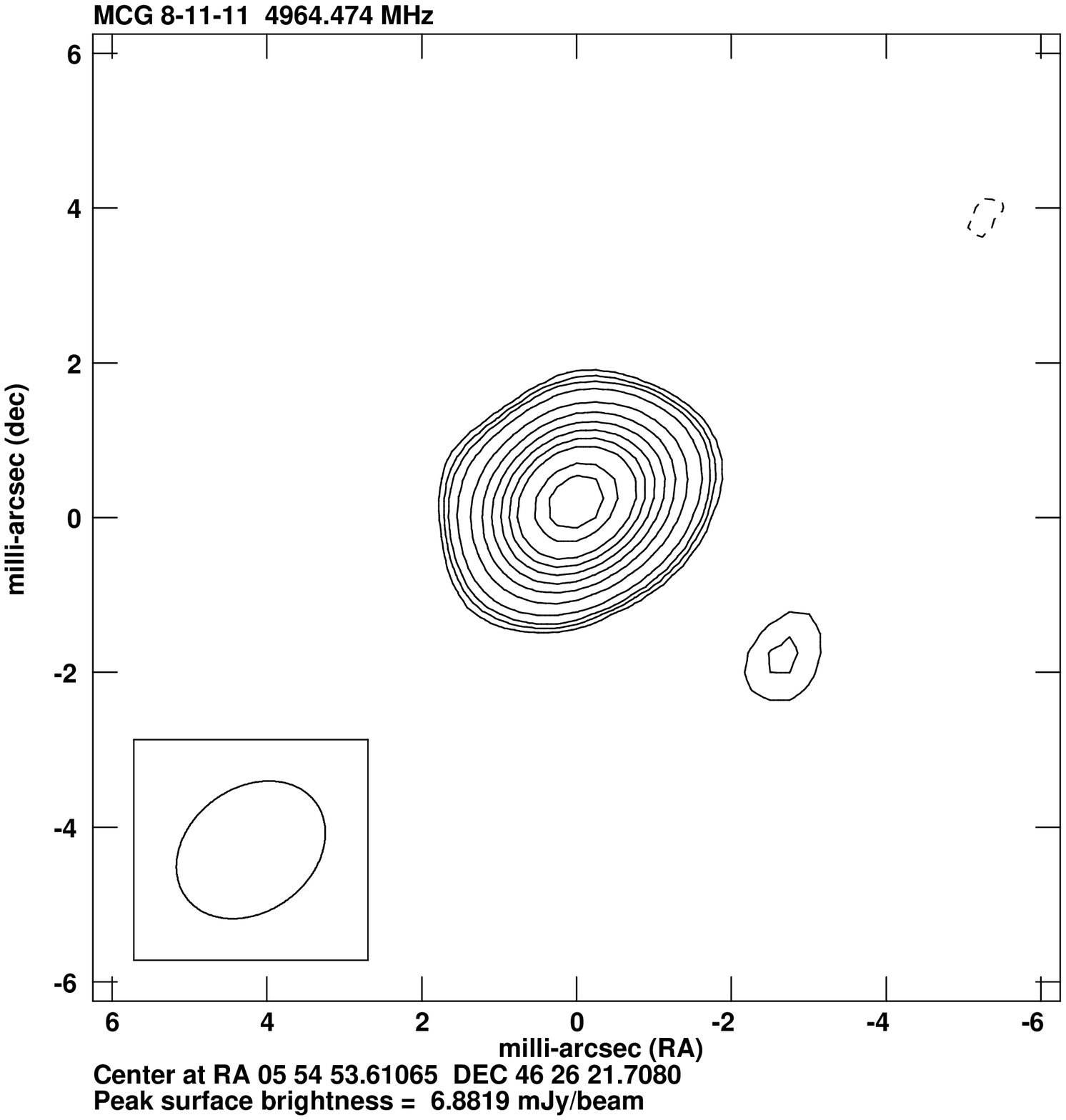,height=2.2in} }
\hbox{
\psfig{figure=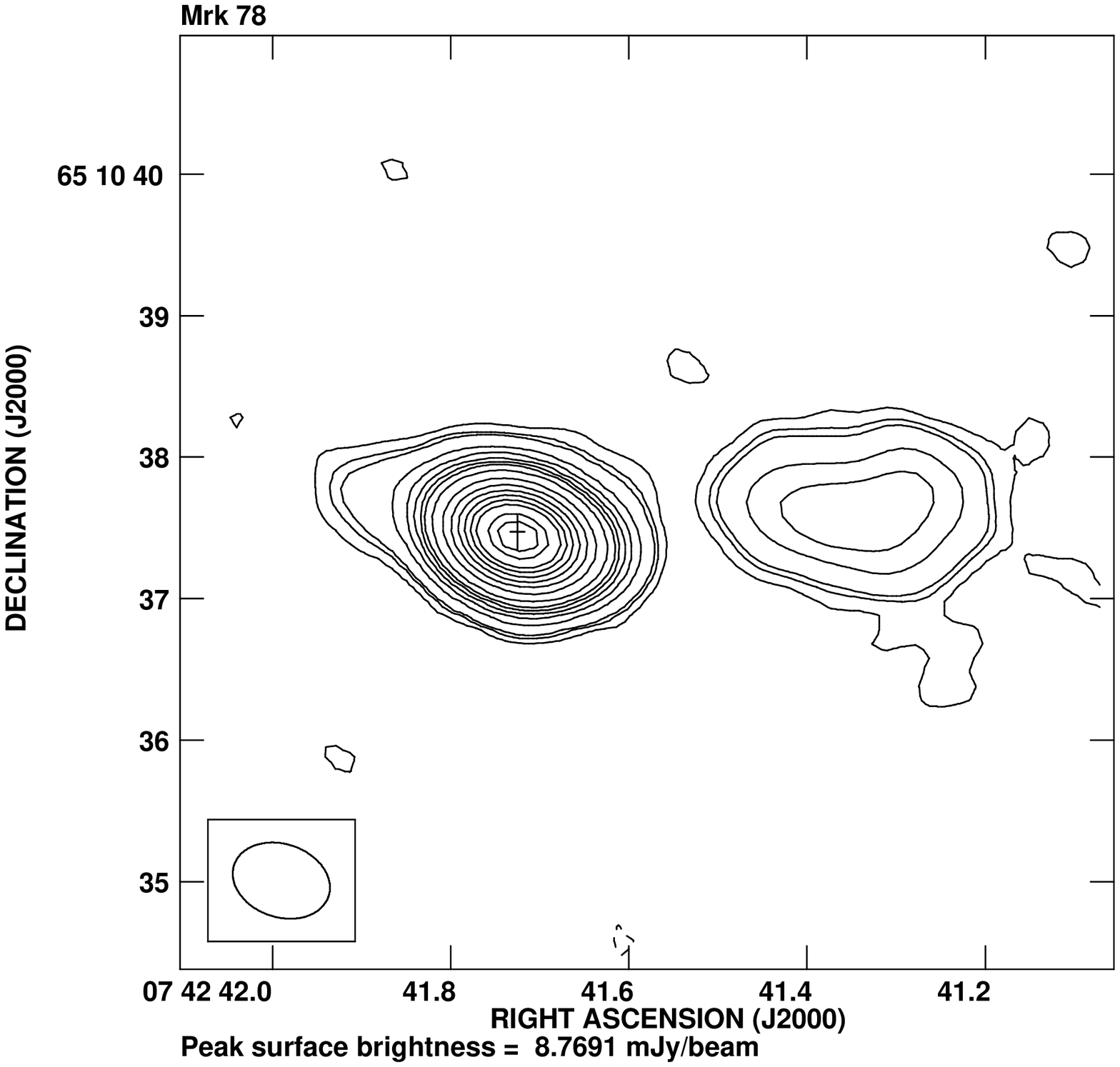,width=2.2in}
\hspace{1.0cm}
\psfig{figure=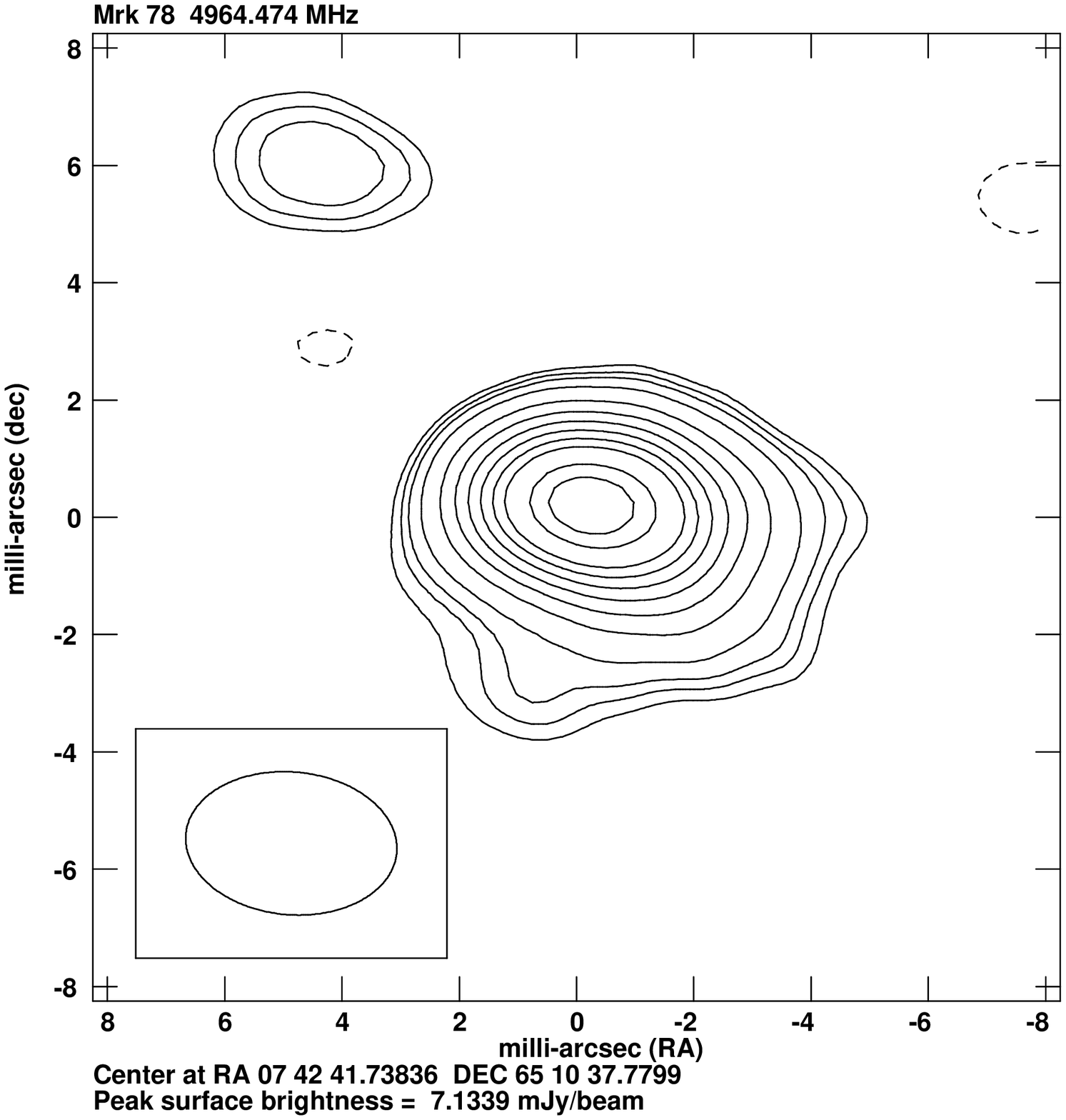,height=2.2in} }
\end{center}
\caption{\footnotesize
{6~cm {\small VLA} ({\it Upper, left panel}; the contour levels are at
--0.6, 0.6, 1, 2, 4, 6, 8, 10, 12, 16, 24, 32, 40, 48, 56, 64, 80 \& 90\%
of the peak surface brightness) \& {\small VLBI}
({\it Upper, right panel}; the contour levels are at --8, 8, 10, 12, 16,
24, 32, 40, 48, 56, 64, 80 \& 90\% of the peak surface brightness)
image of MCG~8-11-11 and}
{6~cm {\small VLA} ({\it Lower, left panel}; the contour levels are at
--1, 1, 1.6, 2, 4, 6, 8, 10, 12, 16, 24, 32, 40, 48, 56, 64, 80 \& 90\%
of the peak surface brightness) \& {\small VLBI}
({\it Lower, right panel}; the contour levels are at --8, 8, 10, 12, 16,
24, 32, 40, 48, 56, 64, 80 \& 90\% of the peak surface brightness)
image of Mrk~78. The small ellipse at the
corner of the image gives the size ({\small FWHM}) of the beam.} }
\label{fig2}
\end{figure}

We detected all 15 of our observed objects on both arcsecond
and milliarcsec~scales. Figure~\ref{fig1} shows the VLA and VLBI
images for two of the sample objects, NGC~2639 (Sy~1) and Mrk~533
(Sy~2). Figure~\ref{fig2} shows examples of MCG~8-11-11 (Sy~1) and
Mrk~78 (Sy~2).
\noindent The Figures also show that we detect either a single component
or multiple components on parsec~scales for our sample sources.
We have performed a Mann-Whitney~U test (Siegal \& Castellan 1988)
and conclude that:
\begin{enumerate}
\item [(i)] a starburst alone cannot power these radio sources because,
(a) they have
high brightness temperature, and (b) the core radio luminosity
at 5~GHz is $\sim$ 10$^{28}$~ergs~s$^{-1}$Hz$^{-1}$
and it arises from a region smaller than a few cubic~parsec.
\item [(ii)] the distributions of radio luminosities on parsec~scales
for the two classes of Sy~galaxies show no statistically significant
difference at a significance level of 0.05 (and also on
kiloparsec~scales at a significance level of 0.10),
\item [(iii)] the fraction of radio emission detected on milliarcsec~scales
({\it i.e.} total radio emission detected with VLBI) to the
emission detected on kiloparsec~scales ({\it i.e.}
total VLA radio emission) is not significantly different
for the two Sy~sub-classes at a significance level of 0.10
and
\item [(iv)] the ratio of compact radio emission ({\it i.e.} emission detected
with VLBI) to the extended radio emission detected on kiloparsec~scales
({\it i.e.} total radio emission~minus~the core radio emission detected
with VLA), using our simultaneous VLBI and VLA measurements is also not
significantly different for the two Sy~sub-classes even at a level of
0.10.  If the jets were significantly relativistically beamed, we would
expect Sy~1s to show systematically more prominent compact radio
emission than Sy~2s.
\end{enumerate}

\smallskip
We find that Sy~1 and Sy~2 galaxies have equal tendency to show compact
radio structures and our results do not agree with the results obtained
by Roy {\it et~al.} (1994) (based on the significantly different
detection rates of Sy~1 and Sy~2 galaxies, Roy {\it et~al.} (1994)
concluded that compact radio structure are much more common in
Sy~2 than in Sy~1 galaxies).

\smallskip
Using the 12, 25, 60 and 100~$\mu$ms flux densitie measurements
using IRAS for our
Sy~sample, we find that Sy~galaxies have similar distribution of
infrared power at all IRAS wavebands at a significance level better
than 0.10.  While the mid and far-infrared flux densities correlate
with the kiloparsec~scale radio flux densities, they do not
correlate with the parsec~scale flux densities. This suggests that
the infrared emission is dominated by non-nuclear emission and a
significant contribution to the infrared emission could be due to
the galaxy disk.

\smallskip
Using the X-ray luminosity measurements (data are available
for 7 out of 10 Sy~1s and 9 out of 10 Sy~2s)
and photon indices (data are available for 9 out of
10 Sy~1s and 6 out of 10 Sy~2s)
from the literature, we test the unification scheme
in this waveband as well.
The Mann-Whitney~U test shows that the distributions
of X-ray (2--10~keV) luminosity for the two classes of Seyfert
galaxies are significantly different (the probability, $P$
being different is less than 0.02).
Seyfert~1 galaxies have systematically higher X-ray luminosity than
Seyfert~2 galaxies. Our result is consistent
with the unified scheme hypothesis. The unified scheme predicts
higher luminosity for Seyfert~1 galaxies than for
Seyfert~2 galaxies because the X-ray emission
in Seyfert~2 galaxies gets attenuated due to the presence of
the obscuring torus in our line of sight whereas the X-rays from
Seyfert~1 galaxies do not suffer from such obscuration.
We also find that the Mann-Whitney~U test shows
that the distributions of 2--10~keV photon index for the two 
classes of Seyfert galaxies are not statistically different even
at a significance level of 0.10.
This result is also consistent with the unified scheme, because we
do not expect flattening of the X-ray spectrum in the hard X-ray
regime for Sy~2 galaxies due to the presence of obscuring torus
as is seen in the soft X-ray regime (Kruper {\it et~al.} (1990).

\section{Conclusions}

In the thesis we have presented radio images of 15 objects that were
observed by us with VLBI and 4 objects that had
VLBI images in the literature. The maps presented in
the Chapter~3 of the thesis reveal, for the first time, the
milliarcsec~scale radio structures of a large carefully selected sample of
Sy galaxies. These measurements provide
a useful database for investigating the generic properties of compact
radio cores in the nuclei of Sy galaxies.

\smallskip
We therefore conclude that Sy~1 and Sy~2 galaxies have equal tendency
to show compact radio structures, in contrast to the results of
Roy {\it et~al.} (1994),~who~concluded that compact radio structures
were much more common in Sy~1s than in Sy~2s. Our results so far appear
to be consistent with the unification scheme hypothesis: the radio
compactnesses of the Sy~1s and Sy~2s are similar.

\section*{Acknowledgements}
I wish to express my thanks to my supervisor Dr. Prajval Shastri 
and our collaborator Dr.~Denise Gabuzda for
their constant guidance and support provided to me as a Ph.D. student.
I thank IIA and IISc (JAP, Phy Dept)  institutes for providing me with the
necessary facilities and support for my work.
I thank the JIVE-institute for their support during my stay at JIVE.
This thesis work involves observations made on NRAO and EVN
instruments. It has also got partial support from the "Indo-Russian
Integrated Long Term Programme."
This research has made use of the ADS, NED and SIMBAD databases.

\label{lastpage}

\begin{thebibliography}{100}

\bibitem{AM89} Antonucci, R.R.J., \& Miller, J.S., 1985, 
  \newblock {\it Astrophys. J.}, {\bf 297}, 621.
\bibitem{DD88} Dahari, O., \& De~Robertis, M.M., 1988, 
  \newblock {\it Astrophys. J. Suppl.}, {\bf 67}, 249.
\bibitem{KW74} Khachikian, E.Y., \& Weedman, D.W., 1974,
  \newblock {\it Astrophys. J.}, {\bf 192}, 581.
\bibitem{Kr90} Kruper, J.S., Canizares, C.R., \& Urry, C.M., 1990,
  \newblock {\it Astrophys. J. Suppl.}, {\bf 74}, 347.
\bibitem{Pr99} Pringle, J.E., Antonucci, R.R.J., Clarke,
C.J., Kinney, A.L., Schmitt, H.R., \& Ulvestad, J.S., 1999,
  \newblock {\it Astrophys. J. Lett.}, {\bf 526}, L9.
\bibitem{NM99} Nagar, N.M., \& Wilson, A.S., 1999, 
  \newblock {\it Astrophys. J.}, {\bf 516}, 79.
\bibitem{Na99} Nagar, N.M., Wilson, A.S., Mulchaey, J.S.,
\& Gallimore, J.F., 1999, 
  \newblock {\it Astrophys. J. Suppl.}, {\bf 120}, 209.
\bibitem{Roy94} Roy, A.L., Norris, R.P., Kesteven, M.J., Troup, E.R.,
\& Reynolds, J.E., 1994, 
  \newblock {\it Astrophys. J.}, {\bf 432}, 496.
\bibitem{Sie88} Siegal S, \& Castellan Jr. N.J., 1988, 
  \newblock {\it Nonparametric statistics for the Behavioral Sciences}, \\
             McGraw Hill Book Company New York.

\end{thebibliography}
\end{document}